\documentclass[groupedaddress,showpacs,showkeys,amssymb,eqsecnum,aps]{revtex4}     
\usepackage[pdftex]{graphicx}
\usepackage{amsmath} 
\usepackage{epsf}                                                                                           
\usepackage{color}                   
\usepackage{verbatim}                                                                         

\newcommand{\be}{\begin{equation}}

\newcommand{\ee}{\end{equation}}

\newcommand{\tr}{{\rm Tr\,}}

\begin{document}                                                                                              
                                                                                                              
\title{First order formulation of the Yang-Mills theory in a background field}

\author{F. T. Brandt and J. Frenkel}
\email{fbrandt@usp.br, jfrenkel@if.usp.br}
\affiliation{Instituto de F\'{\i}sica, Universidade de S\~ao Paulo, S\~ao Paulo, SP 05508-090, Brazil}
\author{D. G. C. McKeon}
\email{dgmckeo2@uwo.ca}
\affiliation{
Department of Applied Mathematics, The University of Western Ontario, London, ON N6A 5B7, Canada}
\affiliation{Department of Mathematics and Computer Science, Algoma University,
Sault Ste. Marie, ON P6A 2G4, Canada}
                                                                                                              
\date{\today}
                                                                                                              
\begin{abstract}                                                                                              
The background gauge renormalization of the first order formulation of the Yang-Mills theory
is studied by using the BRST identities.
Together with the background symmetry, these identities allow for an iterative proof of renormalizability to
all orders in perturbation theory.
However, due to the fact that certain improper diagrams
which violate the  BRST symmetry should be removed, the renormalizability must be 
deduced  indirectly. 
The recursive method involves rescalings and mixings of
the fields, which lead to a  renormalized effective action for the background field theory.  
\end{abstract}                                                                                                
                  

\maketitle

\section{Introduction}
Computing higher order radiative corrections in Yang-Mills theory is currently of great importance in order
to test phenomenological models using precision experiments. Usually these computations are done using the
second order form of Yang-Mills theory with its complicated three and four-point vertices. Background field
quantization has long been known to streamline such calculations as gauge invariance in the background field
is unbroken by gauge fixing \cite{Dewitt:1967ub,KlubergStern:1974xv,Abbott:1980hw,abbott82,Abbott:1983zw,McKeon:1994ds,Grassi:1995wr,weinberg:book95,Barvinsky:2017zlx,Frenkel:2018xup,Lavrov:2018thn}. 
What is less appreciated is that many of these calculations become more compact if the first order formalism
for Yang-Mills theory is
used \cite{Martellini:1997mu,Andrasi:2007dk,Brandt:2015nxa,Brandt:2016eaj,Frenkel:2017xvm,Arnowitt:1962hi,Brandt:2018avq}; this is because in this case there is only one  simple cubic vertex involving the gauge fields
and this vertex is independent of momentum. In this paper we demonstrate how to renormalize the first order
form of the Yang-Mills field when background field quantization is used. Some interesting and unexpected
subtleties arise when doing this. After providing a general analysis of this problem we illustrate our
results by a calculation of the two point function for the background field 
using the first order formalism, which leads to the correct result for
the $\beta$-function.

The background field
method \cite{Dewitt:1967ub,KlubergStern:1974xv,Abbott:1980hw,abbott82,Abbott:1983zw,McKeon:1994ds,Grassi:1995wr,weinberg:book95,Barvinsky:2017zlx,Frenkel:2018xup,Lavrov:2018thn} 
is a formulation which allows to fix a gauge and evaluate the quantum corrections 
without breaking the background gauge symmetry. This is an efficient
method for calculating the    $\beta$-function and it has also been used in perturbative gravity  \cite{tHooft:1973bhk,Goroff:1986th},
The main idea of this method is to write the gauge field   $A^{a}_\mu$ which occurs in the Yang-Mills (YM) Lagrangian 
\be\label{eq1}
{\cal L}_{YM}^{(A)} = -\frac 1 4 \left(\partial_\mu A^{a}_\nu  - \partial_\nu A^{a}_\mu  + g f^{abc} A^{ b}_\mu A^{ c}_\nu \right)^2
\equiv -\frac 1 4 \left(f_{\mu\nu}^a(A)\right)^2
\ee
as $B^{a}_\mu  + Q^{a}_\mu$, where $B^{a}_\mu$   is a background field and $Q^{a}_\mu$    is the quantum field.
Then a gauge is chosen which suppresses the gauge invariance of the $Q^a_\mu$     field, but maintains the gauge invariance in terms
of the  $B^{a}_\mu$    field.
The gauge-fixing term is made dependent upon $B^{a}_\mu$ as
\be\label{eq2}
         {\cal L}_{GF} = -\frac{1}{2\xi}\left[\left(\partial_\mu\delta^{ab} + g f^{acb} B_\mu^c \right) Q^{b\mu}\right]^2
         \equiv - \frac{1}{2\xi}\left(D_\mu^{ab}(B)  Q^{b\mu}\right)^2
\ee
where $\xi$ is a gauge fixing parameter and $D_\mu^{ab}(B)$ is the co-variant derivative.

The above terms are invariant under the 
gauge transformations
\be\label{eq3}
\delta B_\mu^{a}=D_\mu^{ab}(B) \omega^b(x);\;\;\delta Q_\mu^{a}=gf^{abc} Q_\mu^b \omega^c(x);\;\;
\ee 
where $\omega^c(x)$ is an arbitrary infinitesimal parameter. 

Vertices involving $Q$-fields are used inside diagrams while vertices involving $B$-fields
are used for external lines. There are only $Q$-propagators since the $B$-field invariance
has not been broken. Thus, quantum calculations of Green's functions can be performed and
explicit gauge invariance in the background field is maintained. To evaluate the effective
action which generates these functions to one-loop order, no source $J$ is needed (although this
is necessary at higher loops as discussed below). This may be done by carrying out
the path integral over $Q$ in the Lagrangian ${\cal L}(B,Q)$. But, by shifting $Q$ back to $A$,
this would give a trivial result unless the graphs resulting from vertices which are linear in
$Q$ are omitted \cite{weinberg:book95}. In general, the effective action calculated by keeping
all vertices would not be the appropriate one for the background field because this leads
to unwanted one-particle reducible (1PR) graphs. As has been argued in
\cite{Abbott:1980hw,abbott82,Frenkel:2018xup}, in order to get the proper effective action
which generates the one-particle irreducible (1PI) diagrams, it is necessary to
omit the graphs resulting from vertices which are linear in $Q$. This may be implemented
by subtracting from ${\cal L}(B+Q)$ in \eqref{eq1}, the terms linear in $Q$. We point out
that the omission from the effective action of vertices with only one outgoing $Q$ line is
also required in order to obtain the correct S-matrix in the background gauge \cite{Abbott:1983zw}.


When the background method is used to two-loop order or higher,
the sub-graphs are functionals of $B^a_\mu$ as well
as of $Q^a_\mu$.
We may introduce a current $J_\mu^a$ interacting via $J^a_\mu Q^{a\mu}$, thus defining
a functional $Z[B,J]$ which yields a generating functional $W[B,J]=-i\ln Z[B,J]$
for the connected Green functions. This leads, by a Legendre transformation,
to an action   $\Gamma[g,B,Q]$ which has a background symmetry under \eqref{eq3}.
This symmetry is not sufficient to fix the renormalization of  $\Gamma[g, B, Q]$. In addition, one must also
use the BRST symmetry   \cite{brs74} 
of this action.  Although the omission of the linear terms in $Q$  preserves the background symmetry under
\eqref{eq3}, this operation breaks the BRST symmetry. Consequently,
the effective action $\bar\Gamma[g, B, Q]$ is no longer invariant under the BRST transformation.

The background gauge renormalization through the BRST identities has been studied
in a wide class of gauge  theories by Barvinsky et.all \cite{Barvinsky:2017zlx}. But as we have argued,
the exclusion of the contributions linear in the quantum fields, leads 
to a breakdown of the BRST symmetry. This fact poses a problem in
applying the BRST formulation in the background gauge, a point which
has not yet been addressed in the general work of \cite{Barvinsky:2017zlx}. The
outcomes of this relevant issue for renormalizability have been
examined in \cite{Frenkel:2018xup}, in the second order formalism. 
The main purpose of the present paper is to extend this analysis to
the first order formalism, which might further clarify the reasoning behind  
the use of the BRST symmetry in the background gauge. 

Calculations of quantum corrections in the  standard second-order YM theory are generally involved,
due to the presence in \eqref{eq1} of momentum-dependent three-point as well as four-point vertices.
It is well known   \cite{Martellini:1997mu, Andrasi:2007dk, Brandt:2015nxa, Brandt:2016eaj, Frenkel:2017xvm,Arnowitt:1962hi, Brandt:2018avq} 
that   one may replace \eqref{eq1} by a simpler first order Lagrangian, provided one introduces in the theory
another auxiliary field  $F^a_{\mu\nu}$.   The corresponding first order Lagrangian may  then be written as (see section 2)
\be\label{eq4}
         {\cal L}^{I}_{YM}(A,F) = \frac 1 4 F_{\mu\nu}^{ a} F^{\mu\nu\, a}
         -\frac 1 2 F^a_{\mu\nu} f^{a\,\mu\nu}(A).
\ee
This simplifies the computations since the interaction term involves only a single cubic
vertex $\langle FAA \rangle$ which is momentum-independent.
(This formulation is also useful in quantum gravity, where it allows to
replace complicated multiple graviton couplings by simple cubic vertices .)
We now substitute in \eqref{eq4}  $A^a_\mu$ by $B^a_\mu+Q^a_\mu$ and introduce a classical
background ${\cal F}_{\mu\nu}^a$ for the quantum field $F^a_{\mu\nu}$. 
Proceeding along the lines indicated above,
one gets  an action  $\Gamma[g, B, Q, {\cal F}, F]$
which is BRST invariant and has a background symmetry under  
\be\label{eq5}
\delta B_\mu^{a}=D_\mu^{ab}(B) \,\omega^b(x);\;\;\delta Q_\mu^{a}=gf^{abc} Q_\mu^b \,\omega^c(x);\;\;
\delta {\cal F}_{\mu\nu}^{a}=gf^{abc}{\cal F}_{\mu\nu}^b \,\omega^c(x);\;\;
\delta F_{\mu\nu}^{a}=gf^{abc}F_{\mu\nu}^b \,\omega^c(x).
\ee 
These symmetries are sufficient to ensure the renormalizability
of $\Gamma[g, B, Q, {\cal F}, F]$.
But the subtraction of the terms linear in $Q$, which
leads to the correct effective action
$\bar\Gamma[g, B, Q, {\cal F}, F]$
breaks the BRST symmetry.
Nevertheless, as discussed in secs.   3 and 4, the renormalizability of $\Gamma[g, B, Q, {\cal F}_, F]$
can be used in an indirect way to renormalize to all orders the effective action $\bar\Gamma[g, B, Q, {\cal F}, F]$,
which generates the one-particle irreducible graphs. 
To this end, we employ a similar approach to that used in \cite{Frenkel:2018xup} for the second-order YM theory.          
In Sec. 5, we give a short discussion of the results  and point out a possible application of this method 
to the quantum gauge theory  of gravity. 
In Appendix A, a functional equation for the one-particle irreducible generating functional is derived.  In Appendices 
B and C we explicitly show, by computing the $BB$ self-energy, that one must remove the terms linear in $Q$ in order 
to obtain the correct $\beta$-function which leads to asymptotic freedom.      
In Appendix D we discuss the renormalization of the background field
${\cal F}^a_{\mu\nu}$ which, similarly to the quantum field $F^a_{\mu\nu}$, involves
rescalings and mixings.

\section{The first order background field formulation}
In going over from the second order to the first order formulation,
one must ensure that these approaches are equivalent \cite{Martellini:1997mu,Frenkel:2017xvm}
in the sense that they lead to the same Green functions when all external lines are $B$-fields.
Our procedure preserves this condition, which is also attested by the fact that we obtain
the same gauge-invariant result for the $BB$ self-energy. 

The generating functional of Green functions in the second order background formalism is
\be\label{eq6}
Z[B,J]= N \int {\cal D} Q {\cal D} c {\cal D} \bar c
\exp i \int d^4x\left\{{\cal L}_{YM}(B+Q) - \frac{1}{2\xi}\left[D_\mu(B)  Q^{\mu}\right]^2
-\left[\bar c D_\mu(B)\right]\cdot \left[D^\mu(B+Q)    c\right] + J_\mu \cdot Q^\mu
\right\}
\ee
where $c$ and $\bar c$ are ghost fields and we have suppressed the colour indices 
by using the notation $B_\mu\cdot Q_\nu\equiv B^a_\mu Q_\nu^a$ and
$(B_\mu\wedge Q_\nu)^a\equiv f^{abc} B^b_\mu Q_\nu^c$.
To convert \eqref{eq6} to the first order form, we introduce in the normalization constant $N$ the factor
\be\label{eq7}
\int {\cal D} F \exp \frac i 4 \int d^4 x F_{\mu\nu} \cdot F^{\mu\nu} = 
\int {\cal D} F \exp \frac i 4 \int d^4 x \left[F_{\mu\nu}
-f_{\mu\nu}(B+Q)
\right]^2
\ee
where we made a shift in the integration variable. 
Substituting this factor in \eqref{eq6}, leads to the cancellation of ${\cal L}_{YM}(B+Q)$,
with the result
\begin{eqnarray}\label{eq8}
Z[B,{\cal F},J,K]&=& N \int {\cal D} Q {\cal D} c {\cal D} \bar c {\cal D} F
\exp i \int d^4x\left\{\frac 1 4 {\cal F}_{\mu\nu}\cdot {\cal F}^{\mu\nu} +\frac 1 4 F_{\mu\nu}\cdot F^{\mu\nu} 
- \frac 1 2 F^{\mu\nu}\cdot f_{\mu\nu}(B+Q)
\right .
\nonumber \\ && \left .
- \frac{1}{2\xi}\left[D_\mu(B)  Q^{\mu}\right]^2
-\left[\bar c D_\mu(B)\right]\cdot \left[D^\mu(B+Q)   c\right]
+ J_\mu \cdot Q^\mu + K_{\mu\nu} \cdot F^{\mu\nu},
\right\}.
\end{eqnarray}
where we have introduced a gauge invariant background term ${\cal L}_{YM}({\cal F})$ and a
source $K_{\mu\nu}$ for the field $F^{\mu\nu}$.
This leads to a generating functional
for connected Green functions $W[B,{\cal F},J,K] = -i\log Z[B,{\cal F},J,K]$ and hence,
by a Legendre transform 
an action $\Gamma[g,B,Q,{\cal F},F]$.
But this action is not the appropriate one for the background method,                   
since it contains some 1P-reducible graphs.  This shortcoming may be avoided by subtracting
from ${\cal L}(B+Q)$ in \eqref{eq6}, the terms linear in $Q$, which
yields the generating functional in the second order formalism
\begin{eqnarray}\label{eq9}
\bar Z[B,J] &=& N \int {\cal D} Q {\cal D} c {\cal D} \bar c
\exp i \int d^4x\left\{{\cal L}_{YM}(B+Q) - 
Q_\nu \cdot D_\mu(B) f^{\mu\nu}(B) \right.
-{\cal L}_{YM}(B)
\nonumber \\ && \left.
- \frac{1}{2\xi}\left[D_\mu(B)  Q^{\mu}\right]^2
-\left[\bar c D_\mu(B)\right]\cdot \left[D^\mu(B+Q)   c\right] 
+ J_\mu \cdot Q^\mu
\right\},
\end{eqnarray}
where we have subtracted also the  ${\cal L}_{YM}(B)$ term, which is not relevant for our purposes.
To convert \eqref{eq9} into a first order form, 
it is convenient to introduce in N the factor (compare with \eqref{eq7})
\be\label{eq10}
\int {\cal D} F \exp \frac i 4 \int d^4 x \left\{F_{\mu\nu}
-\left[\partial_\mu Q_\nu-\partial_\nu Q_\mu +  g \left(Q_\mu \wedge Q_\nu 
+ Q_\mu \wedge B_\nu + B_\mu \wedge Q_\nu  \right)\right]
\right\}^2 .
\ee
Here it is useful to make a shift
$F_{\mu\nu}\rightarrow F_{\mu\nu} + {\cal F}_{\mu\nu}$,
where ${\cal F}_{\mu\nu}$ is a background for the quantum field ${F}_{\mu\nu}$.
This yields a non-trivial result when one subtracts the term
${\cal F}_{\mu\nu} D^\mu(B) Q^\nu$, which is linear in $Q$.
Such a subtraction preserves the background gauge-invariance
under \eqref{eq5}. Then, substituting the corresponding
expression obtained from \eqref{eq10} into \eqref{eq9}
leads to several cancellations, giving
\begin{eqnarray}\label{eq11}
\bar Z^\prime[B,{\cal F},J] &=& N \int {\cal D} Q {\cal D} c {\cal D} \bar c {\cal D} F
\exp i \int d^4x\left\{
\frac 1 4 {\cal F}_{\mu\nu} \cdot {\cal F}^{\mu\nu} +
\frac 1 2 {   F}_{\mu\nu} \cdot {\cal F}^{\mu\nu} +
\frac 1 4 F_{\mu\nu} \cdot F^{\mu\nu} \right.
\nonumber \\ && \left.
- \frac 1 2 F^{\mu\nu}  \cdot \left[f_{\mu\nu}(Q) + g (B_\mu\wedge Q_\nu+Q_\mu\wedge B_\nu )\right]\right.
-\frac 1 2 g (f_{\mu\nu}(B)+{\cal F}_{\mu\nu}) \cdot (Q^\mu\wedge Q^\nu) 
\nonumber \\ && \left.
- \frac{1}{2\xi}\left[D_\mu(B)  Q^{\mu}\right]^2
-\left[\bar c D_\mu(B)\right]\cdot \left[D^\mu(B+Q)  c\right]
+ J_\mu \cdot Q^\mu  
\right\}.
\end{eqnarray}
We note that in \eqref{eq11}, the bilinear terms
${\cal F} F$ , ${ F} Q$ and $Q Q$  could lead to a mixed
matrix-propagator involving the fields ${\cal F}$, $F$ 
and $Q$. But in order to get a proper background field
${\cal F}$ that occurs only in external lines, this mixing must
be avoided, which may be ensured by removing the bilinear
term ${\cal F}^{\mu\nu} F_{\mu\nu}$. This omission is allowed
since it maintains the background gauge invariance under
\eqref{eq5}. Thus, introducing a source $K_{\mu\nu}$, we get
for the appropriate generating functional in the
first order formulation, the result
\begin{eqnarray}\label{eq11n}
\bar Z[B,{\cal F},J,K] &=& N \int {\cal D} Q {\cal D} c {\cal D} \bar c {\cal D} F
\exp i \int d^4x\left\{
\frac 1 4 {\cal F}_{\mu\nu} \cdot {\cal F}^{\mu\nu} +
\frac 1 4 F_{\mu\nu} \cdot F^{\mu\nu} \right.
\nonumber \\ && \left.
- \frac 1 2 F^{\mu\nu}  \cdot \left[f_{\mu\nu}(Q) + g (B_\mu\wedge Q_\nu+Q_\mu\wedge B_\nu )\right]\right.
-\frac 1 2 g (f_{\mu\nu}(B)+{\cal F}_{\mu\nu}) \cdot (Q^\mu\wedge Q^\nu) 
\nonumber \\ && \left.
- \frac{1}{2\xi}\left[D_\mu(B)  Q^{\mu}\right]^2
-\left[\bar c D_\mu(B)\right]\cdot \left[D^\mu(B+Q)  c\right]
+ J_\mu \cdot Q^\mu  + K_{\mu\nu} \cdot F^{\mu\nu}
\right\}.
\end{eqnarray}

Using the generating functional
$\bar W[B,{\cal F},J,K] = -i\ln\bar Z[B,{\cal F},J,K]$ 
and making a Legendre transform, leads to the correct
effective action $\bar \Gamma[g,B,Q,{\cal F},F]$ in the background gauge.  Let us now compare the Lagrangians which appear in the exponentials of Eqs. \eqref{eq8}  and \eqref{eq11}.
We get, setting $J=K=0$, respectively
\be\label{eq13}
{\cal L}^\prime(g,B,Q,{\cal F},F) =
\frac 1 4 {\cal F}_{\mu\nu}\cdot {\cal F}^{\mu\nu}+
\frac 1 4 F_{\mu\nu}\cdot F^{\mu\nu}
- \frac 1 2 F^{\mu\nu}\cdot f_{\mu\nu}(B+Q)
- \frac{1}{2\xi}\left[D_\mu(B)  Q^{\mu}\right]^2
-\left[\bar c D_\mu(B)\right]\cdot \left[D^\mu(B+Q)   c\right]
\ee
and
\be\label{eq14}
\bar{\cal L}^\prime(g,B,Q,{\cal F},F)
= {\cal L}^\prime(g,B,Q,{\cal F},F)
+\frac 1 2 F_{\mu\nu} \cdot f^{\mu\nu}(B)
- \frac 1 2 g (Q_\mu\wedge Q_\nu)
\cdot ({\cal F}^{\mu\nu} + f^{\mu\nu}(B)).
\ee
The difference between these Lagrangians involves terms which are invariant under the
background transformation \eqref{eq5}.
The second contribution on the right side of \eqref{eq14} subtracts from
${\cal L}^\prime$ a term which would lead to 1P-reducible graphs.  
Moreover, the last term in \eqref{eq14},
which is induced by the subtraction of linear terms in
$Q$, 
is also necessary to obtain correct physical results.
For example, using the Feynman rules derived
in Appendix B, we have evaluated in Appendix C the divergent
part of the background field self-energy.
In a space-time of dimension $d=4-2\epsilon$,  we get to one-loop order  
\be\label{eq15}
\left.\Pi_{\mu\nu}^{ab}(k)\right|_{UV} = -\frac{11}{3} i \delta^{ab} \frac{N g^2}{16\pi^2\epsilon}\left(k_\mu k_\nu - k^2\eta_{\mu\nu}\right).
\ee
This transverse form is independent of the gauge-fixing parameter and leads to the expected
result for the $\beta$-function
\be\label{betaF}
\beta = -\frac{11}{3} \frac{N g^3}{16\pi^2}.
\ee
We note here that the unsubtracted Lagrangian \eqref{eq13} would lead instead to a transverse, but gauge dependent self-energy
for the background field (see Appendix C).

\section{The actions $\Gamma$ and $\bar\Gamma$}   
We remark that ${\cal L}^\prime(g,B,Q,{\cal F},F)$
in \eqref{eq13} 
with the gauge-fixing term left out,
is also invariant under the BRST transformations
\be\label{eq16}
\Delta B = 0;\;
\Delta {\cal F}_{\mu\nu} = 0;\;
\Delta Q_\mu = D_\mu(B+Q) c \tau;\; 
\Delta c = -\frac 1 2 g c\wedge c \tau;\; 
\Delta \bar c = 0;\; 
\Delta F_{\mu\nu} = g F_{\mu\nu}\wedge c \tau,
\ee
where $\tau$   is  an infinitesimal anti-commuting constant.  
Let us now add to ${\cal L}^\prime(g,B,Q,{\cal F},F)$
in \eqref{eq13} the Zinn-Justin source terms $U$, $V$, $W$, which are useful for setting up
the BRST equations \cite{ZinnJustin:1974mc} 
and omit the gauge-fixing term \eqref{eq2}.  This leads to the zeroth order action
\be\label{eq17}
\Gamma^{(0)}(g,B,Q,c,\bar c,{\cal F},F;U,V,W) = \int d^4 x {\cal L}
\ee
where $c$, $\bar c$, $U$, $V$, $W$ transform under 
the background transformations \eqref{eq5} in the same way as $Q$, and
\be\label{eq18}
   {\cal L} = \frac 1 4 {\cal F}^{\mu\nu} \cdot {\cal F}_{\mu\nu}+
         \frac 1 4 F^{\mu\nu} \cdot F_{\mu\nu}
         - \frac 1 2 F^{\mu\nu} \cdot f_{\mu\nu}(B+Q)
+ g W^{\mu\nu} \cdot (F_{\mu\nu}\wedge c) + \left[U_\mu + D_\mu(B) \bar c \right]\cdot 
\left[D^\mu(B+Q) c\right] - \frac 1 2 g V \cdot (c\wedge c)
\ee
It may be verified that ${\cal L}$ is invariant under the BRST transformations \eqref{eq16}, provided that the
sources remain unchanged, so that  $\Gamma^{(0)}$  obeys the BRST equations
(where $Q$ actually stands for the mean value of the quantum field)
\be\label{eq19}
\int d^4 x\left[
\frac{\delta \Gamma^{(0)}}{ \delta F_{\mu\nu}}\cdot \frac{\delta \Gamma^{(0)}}{\delta W^{\mu\nu}}+
\frac{\delta \Gamma^{(0)}}{ \delta Q_\mu}\cdot \frac{\delta \Gamma^{(0)}}{\delta U^\mu}+
\frac{\delta \Gamma^{(0)}}{ \delta c}\cdot \frac{\delta \Gamma^{(0)}}{ \delta V}
\right]=0
\ee
\be\label{eq20}
\frac{\delta \Gamma^{(0)}}{ \delta \bar c} - D_\mu(B) \frac{\delta \Gamma^{(0)}}{ \delta U_\mu} = 0.
\ee
Equation \eqref{eq20} is a consequence of the fact that the Lagrangian \eqref{eq18} depends on $U_\mu$        
only through the combination $U_\mu + D_\mu(B) \bar c$.  Moreover, equation \eqref{eq19} may be understood by
rewriting it, with the help of \eqref{eq16}, in the alternative form
\be\label{eq21}
\int d^4 x\left[
\frac{\delta \Gamma^{(0)}}{ \delta F_{\mu\nu}}\cdot \Delta F_{\mu\nu}+
\frac{\delta \Gamma^{(0)}}{ \delta Q_\mu}\cdot \Delta Q_\mu+
\frac{\delta \Gamma^{(0)}}{ \delta c}\cdot \Delta c = 0
\right]
\ee
which reflects the invariance of  $\Gamma^{(0)}$        under the BRST transformations \eqref{eq16}.     

By using an analogous method to that employed in the usual first order
formulation of the YM theory \cite{Frenkel:2017xvm}
one can show that the action $\Gamma$ satisfies to all orders the BRST equation   
\begin{subequations}\label{eq22}
\be\label{eq22a}
\int d^4 x\left[
\frac{\delta \Gamma}{ \delta F_{\mu\nu}}\cdot \frac{\delta \Gamma}{\delta W^{\mu\nu}}+
\frac{\delta \Gamma}{ \delta Q_\mu}\cdot \frac{\delta \Gamma}{\delta U^\mu}+
\frac{\delta \Gamma}{ \delta c}\cdot \frac{\delta \Gamma}{ \delta V}
\right]=0
\ee
\be\label{eq22b}
\frac{\delta \Gamma}{ \delta \bar c} - D_\mu(B) \frac{\delta \Gamma}{ \delta U_\mu} = 0.
\ee
\end{subequations}
The identities resulting from \eqref{eq22} are different from the usual ones
in the Yang-Mills theory due to the dependence on the background field $B(x)$.
We have explicitly verified, to order $g^3$, some examples of these identities.

But as we have explained, $\Gamma$ is not the correct action for the background method.
In order to get the appropriate action $\bar\Gamma$, one must instead start from the Lagrangian \eqref{eq14},
where the last two terms are not BRST invariant,
and use a similar procedure to that which led to the action \eqref{eq17}.
We then find that  $\bar\Gamma^{(0)}$        may be obtained
from   $\Gamma^{(0)}$       by the operation $\Omega^{(0)}$
\be\label{eq23}
\bar\Gamma^{(0)} \equiv \Omega^{(0)}(g,Q,{\cal F},F) \Gamma^{(0)} =
\Gamma^{(0)} +
\int d^4 x \left\{\left[g (\,Q_\mu\wedge Q_\nu)\cdot
  \left(\frac{\delta}{\delta F_{\mu\nu}}
  -\frac{\delta}{\delta {\cal F}_{\mu\nu}}\right)
-F_{\mu\nu}\frac{\delta}{\delta { F}_{\mu\nu}}
  \right]\Gamma^{(0)}\right\}_{F=Q=c=0} .
\ee
In Eq. \eqref{eq23}, only the derivatives should be taken at the particular point $F=Q=c=0$.
Moreover, a $x$-dependence of the fields $Q$,  ${\cal F}$  and $F$  in the operator $\Omega^{(0)}$ is to be understood.  
We note that this operator preserves the background gauge invariance under \eqref{eq5}, but breaks
the BRST symmetry under \eqref{eq16}.
It follows that $\bar\Gamma^{(0)}$ does not satisfy the BRST equations.
The generalization of the above relation, to higher orders, will be examined in the next section.

\section{Renormalization}  
As we pointed out, we study first the renormalization of    $\Gamma$, which requires
both the background invariance as well as the BRST symmetry.
Since the background field $B$ appears explicitly in the BRST Eq. \eqref{eq22b}, we need to fix its renormalization.
To this end we remark that in consequence of the background symmetry under \eqref{eq5}, the renormalized action         
which is got by functionally integrating over $Q$, $c$, $\bar c$, $F$                        and setting the sources to zero,  
must have the form
\be\label{eq26}
\Gamma_R[g,B] = -\frac 1 4 Z_B \int d^4 (\partial_\mu B_\nu - \partial_\nu B_\mu + g B_\mu \wedge B_\nu)^2.
\ee
This may be obtained from the bare action  $\Gamma^{(0)}(g^{(0)},B^{(0)})$, by the re-scalings
\be\label{eq27}
g^{(0)}= Z_g g;\;   B_\mu^{(0)} = Z^{1/2}_B B_\mu = Z^{-1}_g B_\mu .
\ee  
Thus the background invariance ties these two renormalizations by the relation $Z^{1/2}_B Z_g  = 1$,
which is an important virtue of the background field method.  

One must also re-scale in
$\Gamma^{(0)}(g^{(0)},B^{(0)},Q^{(0)},c^{(0)},\bar c^{(0)}, {\cal F}^{(0)},F^{(0)};U^{(0)},V^{(0)},W^{(0)})$
the fields
\be\label{eq28}
Q_\mu^{(0)} = Z_Q^{1/2} Q_\mu;\; c^{(0)} = \tilde Z^{1/2} c;\; \bar c^{(0)} = \tilde Z^{1/2} \bar c.
\ee
As shown in \cite{Frenkel:2017xvm}, the renormalization of the first order formulation of the YM theory
requires a re-scaling as well as a mixing of the $F_{\mu\nu}$ field
\be\label{eq29}
F_{\mu\nu}^{(0)} = Z_F^{1/2} F_{\mu\nu} + Z_{FQ} f_{\mu\nu}(Q)
\ee
where $f_{\mu\nu}$ is defined in \eqref{eq1} and both $Z_F^{1/2}-1$;   $Z_{FQ}$ 
are of order $\hbar$. Similarly, the renormalization of the
background field ${\cal F}_{\mu\nu}$ involves both
a rescaling and a mixing of this field  
\be\label{eq44a}
         {\cal F}^{(0)}_{\mu\nu} =
         Z^{1/2}_{{\cal F}}{\cal F}_{\mu\nu}  +
         Z_{{\cal F} f}{f}_{\mu\nu}(B),  
\ee
where $Z^{1/2}_{{\cal F}}-1$  and $Z_{{\cal F} f}$ 
are gauge dependent quantities of order $\hbar$
(see Appendix D).

In the renormalization process,
the bare sources  $U^{(0)}$, $V^{(0)}$ and $W^{(0)}$
will also undergo appropriate re-scalings and mixings which relate
these to the renormalized sources $U$, $V$ and $W$ \cite{Frenkel:2017xvm}. 
All such transformations preserve  the BRST invariance.  Using
this gauge symmetry together with the Lorentz invariance,  one can show recursively \cite{taylor:1976b, itzykson:1980b}
that the   renormalized action  $\Gamma_R$  must be similar to   $\Gamma^{(0)}$ in Eq. \eqref{eq17},  but it must
include all the allowed re-scalings and mixings.  Thus, it should have the form
\be\label{eq30}
\Gamma_R(g,B,Q,c,\bar c,{\cal F},F;U,V,W) =
\Gamma^{(0)}(g^{(0)},B^{(0)},Q^{(0)},c^{(0)},\bar c^{(0)},{\cal F}^{(0)},F^{(0)};U^{(0)},V^{(0)},W^{(0)}),
\ee
where the bare quantities can be expressed in terms of the renormalized ones as indicated above.

Finally, we must relate the renormalized action $\bar \Gamma_R$ for the background field to $\Gamma_R$.
To this end, we note that all the above transformations preserve as well the background gauge symmetry.
Thus, one may define a renormalized operator by
\be\label{eq4.6}
\Omega_R(g,Q,{\cal F},F) =
\Omega^{(0)}(g^{(0)}, Q^{(0)},{\cal F}^{(0)},F^{(0)})=\Omega^{(0)}[Z_g g; Z^{1/2}_Q Q_\mu;
         Z^{1/2}_{{\cal F}}{\cal F}_{\mu\nu}  +
         Z_{{\cal F} f}{f}_{\mu\nu}(B);  
  Z_F^{1/2} F_{\mu\nu}+Z_{FQ} f_{\mu\nu}(Q)]
\ee
which reduces to lowest order to $\Omega^{(0)}(g,B,F)$ in Eq. \eqref{eq23} and maintains 
to all orders the background gauge invariance.
Hence, to higher orders, the appropriate generalization of Eq. \eqref{eq23} may be written in the form
\be\label{eq30n}
\bar\Gamma_R = \Omega_R(g,Q,{\cal F},F) \Gamma_R
\ee
which allows to deduce the renormalized effective action
$\bar\Gamma_R$ by the application of the operation $\Omega_R$ to $\Gamma_R$.

\section{Conclusion}
Background field quantization has some appealing features, especially when considering
the renormalization of gauge theories. The relation between the coupling constant and the
background field renormalization \eqref{eq27}, has been exploited in explicit calculations in the 
Standard Model \cite{Mckeon:1982kp,Denner:1994xt}. 
In a four dimensional space-time,
this relation leads 
to the condition that 
the divergent part of the background field self-energy should be gauge-independent.
%
In higher dimensions, the YM theory is non-renormalizable and then it is no longer
possible to directly relate $\langle BB\rangle$ to an observable quantity \cite{Brandt:2018kur}. Thus, in this case there is no reason why
the divergent part of the background field self-energy should be gauge independent. All these
features are entirely consistent with the result \eqref{c9} for $\langle BB\rangle$, evaluated
in a general dimension $d$.


The first order formalism for the YM theory has an advantage over the usual second order formalism, in that
the complicated three and four-point vertices of the later are replaced by simple, momentum-independent, cubic vertices
in the former formalism.
An interesting property of this formulation is that the
renormalization of $F_{\mu\nu}$ and its background
${\cal F}_{\mu\nu}$ involves rescalings as well as non-linear mixings
of the fields.

One subtle feature of using the background field method is that the terms linear in the quantum field $Q$ must be
removed. 
This leads to the correct result \eqref{betaF} for the $\beta$-function, but breaks the  BRST symmetry. 
Nevertheless, we have shown 
that the BRST identities can be indirectly used to renormalize the background gauge formulation of the first order YM theory.
To this end, we have first employed the conventional BRST procedure to
renormalize $\Gamma$, and then inferred the renormalization of
$\bar\Gamma$ by implementing the operation $\Omega_R$ 
defined in Eq. \eqref{eq4.6}. Using this method to all orders, we have 
obtained the renormalized effective action \eqref{eq30n}
for the background gauge theory.

The above considerations may hopefully shed some light on how this formulation can be applied to the first order
 (Palatini) form of the Einstein-Hilbert action for General Relativity. Such an action is of particular interest as it involves
only a finite number of interacting cubic vertices \cite{Brandt:2015nxa, Brandt:2016eaj} and allows one to introduce
a graviton propagator that is both traceless and transverse \cite{Brandt:2007td,Brandt:2017oib}.




\begin{acknowledgments}
F. T. B. and J. F. would like to thank CNPq for financial support.
D. G. C. M. would like to thank Roger
Macleod for an enlightening discussion, the Universidade de S\~ao Paulo for the hospitality
during the realization of this work and FAPESP for financial support (Grant Number 2018/01073-5).
We are indebted to J. C. Taylor for helpful discussions.
\end{acknowledgments}

\appendix

\section{Generating function for 1PI Green's Functions}
The necessity of subtracting terms linear in the quantum field $Q^a_\mu$ from the Lagrangian ${\cal L}_{YM}(B+Q)$ when 
computing $\bar Z$ in Eq. \eqref{eq9} can be clarified by considering directly the path integral for the 1PI generating functional 
$\Gamma[\bar Q,B]$ \cite{Culumovic:1989nw}. Rather than working with the generating functional appearing in
Eq. \eqref{eq9} that follows from background field quantization, we consider the 1PI generating
functional that arises with conventional quantization. This generating functional $\hat \Gamma[f]$, which depends on the
average of the quantum field $\phi$, is related  to $\Gamma[\bar Q,B]$ arising in the 
background field method by \cite{Abbott:1980hw,abbott82}
\be\label{a1}
\hat\Gamma[f] = \Gamma[\bar Q =0, B=f]
\ee

If we consider the field $\phi$ with a Lagrangian ${\cal L}(\phi)$, then, in the Euclidean space
\begin{eqnarray}\label{a2}
Z[J] &=& \int {\cal D} \phi \exp \left\{-\frac{1}{\hbar} \int dx[{\cal L}(\phi) + J \phi]\right\} \nonumber \\
&\equiv& \exp \left\{-\frac{1}{\hbar} W[J]\right\}
\end{eqnarray}
leads to a generating functional for 1PI diagrams
\be\label{a3}
\Gamma[f] = W[J] - \int dx f(x) J(x),
\ee
where
\begin{subequations}\label{a4}
\be\label{a4a}
f(x) = \frac{\delta W[J]}{\delta J(x)},
\ee
\be\label{a4b}
J(x) = - \frac{\delta \Gamma[f]}{\delta f(x)}.
\ee
\end{subequations}
Together, Eqs. \eqref{a2} and \eqref{a3} lead to
\be\label{a5a}
\exp \left\{-\frac{1}{\hbar} \Gamma[f]\right\} = \int {\cal D} \phi 
\exp \left\{-\frac{1}{\hbar} \int dx[{\cal L}(\phi) + J(\phi -f)]\right\}
\ee
which, upon making the shift $\phi\rightarrow\phi+f$, becomes
\be\label{a5}
\int {\cal D} \phi 
\exp \left\{-\frac{1}{\hbar} \int dx[{\cal L}(f+\phi) + J\phi ]\right\}.
\ee
In Eq. \eqref{a5}, $J(x)$ is no longer independent as it is in Eq. \eqref{a2};
it is a function of $f(x)$ on account of Eq. \eqref{a4b}.

If we now expand
\be\label{a6}
{\cal L}(f+\phi) = {\cal L}(f) + {\cal L}^\prime(f)\phi + 
\frac{1}{2!}{\cal L}^{\prime\prime}(f)\phi^2+
\frac{1}{3!}{\cal L}^{\prime\prime\prime}(f)\phi^3+
\frac{1}{4!}{\cal L}^{\mbox{iv}}(f)\phi^4
\ee
then Eq. \eqref{a5} becomes
\begin{eqnarray}\label{a7}
\exp \left\{-\frac{1}{\hbar} \Gamma[f]\right\} &=& \exp \left\{-\frac{1}{\hbar} \int dx {\cal L}(x)\right\} 
\exp \left\{-\frac{1}{\hbar} \int dx
\left[ \frac{1}{3!}{\cal L}^{\prime\prime\prime}(f) \left(-\hbar \frac{\delta}{\delta j(x)}\right)^3+
\frac{1}{4!}{\cal L}^{\mbox{iv}}(f) \left(-\hbar \frac{\delta}{\delta j(x)}\right)^4\right]\right\}
\nonumber \\ &&\int {\cal D} \phi 
\exp \left\{-\frac{1}{\hbar} \int dx\left[{\cal L}^{\prime\prime}(f)\phi^2 + j(x)\phi(x) \right]\right\},
\end{eqnarray}
where by Eqs. \eqref{a4} and \eqref{a6}
\be\label{a8}
j(x) = {\cal L}^\prime(f(x)) - \frac{\delta\Gamma[f]}{\delta f(x)}.
\ee

If we now make the loop expansion of $\Gamma[f]$ so that
\be\label{a9}
\Gamma[f] = \Gamma_0[f] +\hbar\Gamma_1[f]+\hbar^2\Gamma_2[f] + \dots
\ee
then upon matching powers of $\hbar$ in Eq. \eqref{a7} we obtain
\be\label{a10}
\Gamma_0[f] = \int dx {\cal L}(f) ,
\ee
\be\label{a11}
\Gamma_1[f] = -\frac 1 2 \log \det {\cal L}^{\prime\prime}(f) ,
\ee
\begin{eqnarray}\label{a12}
\Gamma_2[f] &=& -\frac 1 2 \int dx dy \frac{\delta {\Gamma_1}}{\delta f(x)} \Delta(x-y) \frac{\delta {\Gamma_1}}{\delta f(y)} 
+\frac 1 2 \int dx dy \frac{\delta^3 {\cal L}}{\delta f^3(x)} \Delta(0)\Delta(x-y) \frac{\delta {\Gamma_1}}{\delta f(y)} 
\nonumber \\ &&
-\frac 1 8 \int dx dy \frac{\delta^3 {\cal L}}{\delta f^3(x)} \frac{\delta^3 {\cal L}}{\delta f^3(y)} (\Delta(0))^2\Delta(x-y) 
-\frac{1}{3!2!} \int dx dy \frac{\delta^3 {\cal L}}{\delta f^3(x)} \frac{\delta^3 {\cal L}}{\delta f^3(y)} (\Delta(x-y))^3 
\nonumber \\ &&
+\frac 1 8 \int dx  \frac{\delta^4 {\cal L}}{\delta f^4(x)}  (\Delta(0))^2,
\end{eqnarray}
where 
\be 
\Delta(x-y) =[{\cal L^{\prime\prime}}(f)]^{-1} 
\ee
and by Eq. \eqref{a11}
\be\label{a13}
\frac{\delta\Gamma_1}{\delta f(x)} =  -\frac 1 2 \left( \tr \frac{1}{{\cal L}^{\prime\prime}(f)}  \right)
\frac{\delta^3{\cal L}}{\delta f^3(x)}.
\ee
Upon using \eqref{a13}, we see that $\Gamma_2[f]$ reduces to the last two terms on the right side of Eq. \eqref{a12}
which can be represented graphically by
\[
\begin{picture}(0,0)%
\includegraphics{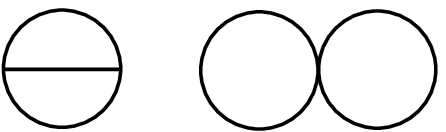}%
\end{picture}%
\setlength{\unitlength}{4144sp}%
\begingroup\makeatletter\ifx\SetFigFont\undefined%
\gdef\SetFigFont#1#2#3#4#5{%
  \reset@font\fontsize{#1}{#2pt}%
  \fontfamily{#3}\fontseries{#4}\fontshape{#5}%
  \selectfont}%
\fi\endgroup%
\begin{picture}(2007,572)(-2705,-438)
\put(-2020,-195){\makebox(0,0)[lb]{\smash{{\SetFigFont{10}{12.0}{\rmdefault}{\mddefault}{\updefault}{\color[rgb]{0,0,0}$+$}%
}}}}
\end{picture}%

\]
which are the two 1PI graphs in background field theory quantization.

In Refs \cite{Abbott:1983zw,Rebhan:1984bg,McKeon:1985uc} it is shown that the $S$-matrix is independent of the
vertices generated by dependence of gauge fixing on the background field, both with covariant gauge fixing
of Eq. \eqref{eq2} and with other non-covariant gauges \cite{Rebhan:1984bg,McKeon:1985uc}.

This procedure can be modified so as to be applied to the generating
functional for the second order formalism Yang-Mills theory,  
justifying the appearance of the term linear in $Q^a_\mu$ 
in Eq. \eqref{eq9} when computing 1PI  graphs from the generating function $Z$. 

\section{Feynman rules}
Here we present the Feynman rules which arise from argument of the exponential $iS$ in \eqref{eq11}.
The bilinear terms in the quantum fields $F$ and $Q$ can be expressed in matrix form as follows 
(here we follow \cite{Brandt:2015nxa})
 \begin{equation}\label{e12}
\frac{1}{2}\left(\begin{matrix}
Q_\mu, & F_{\lambda\sigma}^a
\end{matrix}\right)
\left(\begin{matrix}
  \frac{1}{\xi}\partial^\mu\partial^\nu & \frac{1}{2}\left(\partial^\rho\eta^{\kappa \mu} - \partial^\kappa \eta^{\rho\mu}\right) \\
 -\frac{1}{2}\left(\partial^\lambda\eta^{\sigma \nu} - \partial^\sigma \eta^{\lambda\nu}\right)  & 
\frac{1}{4}\left(\eta^{\lambda\rho}\eta^{\sigma \kappa} - \eta^{\lambda \kappa} \eta^{\sigma\rho}\right) 
\end{matrix}\right)
\left(\begin{matrix}
Q_\nu \\ F_{\rho\kappa}^a
\end{matrix}\right).
\end{equation}
The inverse of the matrix appearing in Eq. \eqref{e12} is
\begin{equation}\label{e13}
\Delta(\partial) = 
\left(\begin{matrix}
\frac{1}{\partial^2}\left(\eta^{\mu\nu} -   \frac{(1-\xi)}{\partial^2}\partial^\mu\partial^\nu\right) 
& -\frac{1}{\partial^2}\left(\partial^\rho\eta^{\kappa \mu} - \partial^\kappa \eta^{\rho\mu}\right) \\
 \frac{1}{\partial^2}\left(\partial^\lambda\eta^{\sigma \nu} - \partial^\sigma \eta^{\lambda\nu}\right)  & 
2\left(I^{\lambda\sigma, \rho \kappa} - \frac{1}{\partial^2}L^{\lambda \sigma,\rho\kappa}\right) 
\end{matrix}\right),
\end{equation}
where
\begin{subequations}\label{e14}
\begin{equation}\label{e14a}
I^{\lambda\sigma,\rho\kappa} = \frac{1}{2}\left(\eta^{\lambda\rho}\eta^{\sigma \kappa} - \eta^{\lambda \kappa} \eta^{\sigma\rho}\right) 
\end{equation}
and
\begin{equation}\label{e14b}
L^{\lambda\sigma,\rho\kappa}(\partial) = \frac{1}{2}\left(\partial^\lambda\partial^\rho \eta^{\sigma\kappa} 
+ \partial^\sigma\partial^\kappa \eta^{\lambda\rho} - \partial^\lambda\partial^\kappa \eta^{\sigma\rho} 
- \partial^\sigma\partial^\rho \eta^{\lambda\kappa} \right). 
\end{equation}
\end{subequations}
From \eqref{e13} we obtain the following expressions for the momentum space propagators of the quantum fields
\be\label{QQprop} 
\begin{array}{lcc}
\input{QQ.pspdftex}   
\end{array} \;\;\;\;  = \displaystyle{\frac{-i\delta^{ab}}{k^2+i0}\left[\eta_{\mu\nu}-(1-\xi)\frac{k_\mu k_\nu}{k^2+i0}\right]},
\ee
\be 
\begin{array}{lcc}
\input{FF.pspdftex} 
\end{array} \;\;\;\; = \displaystyle{i\delta^{ab}
\left[\eta_{\lambda\rho}\eta_{\sigma\kappa} - \eta_{\lambda\kappa}\eta_{\sigma\rho}   
-\frac{1}{k^2+i0} \left(k_\lambda k_\rho \eta_{\sigma\kappa} 
+ k_\sigma k_\kappa \eta_{\lambda\rho} - k_\lambda k_\kappa \eta_{\sigma\rho} 
- k_\sigma k_\rho \eta_{\lambda\kappa} \right)\right]  }    . 
\ee
and
\be 
\begin{array}{lcc}
\input{QF.pspdftex} 
\end{array} \;\;\;\; = \displaystyle{-
 \frac{\delta^{ab}}{k^2+i 0}\left(k_\rho\eta_{\kappa \mu} - k_\kappa \eta_{\rho\mu}\right) }. 
\ee
Similarly, the quadratic term for the ghost fields yields
\be 
\begin{array}{lcc}
\input{cc.pspdftex} 
\end{array} \;\;\;\;  = \displaystyle{
 \frac{i\delta^{ab}}{k^2+i 0}  }. 
\ee

From the interaction terms in \eqref{eq11} we obtain
\be 
\begin{array}{lcc}
\input{FBQ.pspdftex} 
\end{array}  \;\;\; = 
\displaystyle{-\frac{ig}{2} f^{abc} \left(\eta_{\mu\lambda}\eta_{\nu\sigma}-\eta_{\mu\sigma}\eta_{\nu\lambda}\right) },
\ee
\be 
\begin{array}{lcc}
\input{fQQ.pspdftex} 
\end{array}  \;\;\; = 
\displaystyle{-\frac{ig}{2} f^{abc} \left(\eta_{\mu\lambda}\eta_{\nu\sigma}-\eta_{\mu\sigma}\eta_{\nu\lambda}\right) },
\ee
\be 
\begin{array}{lcc}
\input{FQQa.pspdftex} 
\end{array}  \;\;\; = 
\displaystyle{-\frac{ig}{2} f^{abc} \left(\eta_{\mu\lambda}\eta_{\nu\sigma}-\eta_{\mu\sigma}\eta_{\nu\lambda}\right) },
\ee
\be 
\begin{array}{lcc}
\input{BQQ.pspdftex} 
\end{array}  \;\;\; = 
\displaystyle{\frac{g}{\xi} f^{abc} \left(\eta_{\mu\lambda} r_\sigma - \eta_{\mu\sigma} q_\lambda \right) },
\ee
\be\label{b12} 
\begin{array}{lcc}
\input{BBQQ.pspdftex} 
\end{array}  \;\;\; =  
\displaystyle{\frac{-ig^2}{\xi} \left[f^{ace} f^{bde} \eta_{\mu\lambda} \eta_{\nu\rho} 
                                                +f^{ade} f^{bce} \eta_{\mu\rho}\eta_{\nu\lambda} \right]},
\ee
\be\label{b13} 
\begin{array}{lcc}
\input{Bcc.pspdftex} 
\end{array}  \;\;\; = 
\displaystyle{g  f^{abc}  (p+q)_{\mu} },
\ee
\be\label{b14} 
\begin{array}{lcc}
\input{Qcc.pspdftex} 
\end{array}  \;\;\; = 
\displaystyle{g  f^{abc}  p_{\mu} },
\ee
\be\label{b15}  
\begin{array}{lcc}
\input{BQcc.pspdftex} 
\end{array}  \;\;\; = 
\displaystyle{-ig^2  f^{ace} f^{dbe}  \eta_{\mu\nu} },
\ee
\be\label{b16} 
\begin{array}{lcc}
\input{BBcc.pspdftex} 
\end{array}  \;\;\; = 
\displaystyle{-ig^2 \eta_{\mu\nu} \left(f^{ace} f^{dbe} +f^{ade} f^{cbe} \right)},
\ee
where we are using the momentum space representation 
$f_{\mu\nu}^a =i \delta^{ab} \left(k_\mu \eta_{\nu\beta}  -  k_\nu \eta_{\mu\beta}\right) B^{b\, \beta}(k) $.

\section{The $\langle BB \rangle$ self-energy}
The one-loop contributions to the two-point function $\langle BB \rangle$ are given
by the Feynman diagrams of fig. 1 (we are not including tadpole diagrams which arise from the vertices 
\eqref{b12} and \eqref{b16} in the appendix B). 
After the loop momentum integration,  the result can only depend (by covariance) on the two tensors
$\eta_{\mu\nu}$ and $k_\mu k_\nu$. A convenient tensor basis is 
\be
   {\cal T}_{\mu\nu}^1   = k_\mu k_\nu - k^2 \eta_{\mu\nu} \;\; \mbox{and} \;\; {\cal T}_{\mu\nu}^2 = k_\mu k_\nu
\ee
so that each diagram in figure 1 can be written as $\Pi^{I\, ab}_{\mu\nu}(k)  = N g^2 \delta^{ab} \Pi^{I}_{\mu\nu}(k)$
(we are using $f^{amn} f^{bmn} = N \delta^{ab}$), where
\begin{equation}\label{eq2a}
\Pi^{I}_{\mu\nu}(k) =  \sum_{i=1}^{2}  {\cal T} ^i_{\mu\nu}(k) 
C^I_i(k) ; \;\;\; I=\mbox{a, }\mbox{b, }\mbox{c  } \dots \mbox{h} .
\end{equation}
The coefficients $C^I_i$ can be obtained solving the following system of two algebraic equations
\be
\sum_{i=1}^2 {\cal T}^i_{\mu\nu}(k) {\cal T}^j{}^{\mu\nu}(k) C^I_i(k) =
\Pi^I_{\mu\nu}(k) {\cal T}^j{}^{\mu\nu}(k) \equiv J^I{}^j(k); \;\;
j=1,2 .
\ee
Using the Feynman rules for $\Pi^I_{\mu\nu}(k)$ 
the integrals on the right hand side have the following form
\be
J^I{}^j(k) = \int \frac{d^d p}{(2 \pi)^d}  s^I{}^j(p,q,k).
\ee
where $q=p+k$; $p$ is the loop momentum, $k$ is the external momentum and $s^I{}^j(p,q,k)$ are
scalar functions. Using the relations 
\begin{subequations}
\begin{eqnarray}
p\cdot k = (q^2 - p^2 - k^2)/2, \\
q\cdot k = (q^2 + k^2 - p^2)/2, \\
p\cdot q = (p^2 + q^2 - k^2)/2, 
\end{eqnarray}
\end{subequations}
the scalars  $s^I{}^j(p,q,k)$ can be reduced to combinations of powers of $p^2$ and $q^2$. As a result, the
integrals $J^I{}^j(k)$  can be expressed in terms of combinations of the following well known integrals 
\begin{equation}
I^{l m} \equiv 
\int \frac{d^d p}{(2 \pi)^d} \frac{1}{(p^2)^l (q^2)^m} = i \frac{(k^2)^{d/2-l-m}}{(4\pi)^{d/2}}
\frac{\Gamma(l+m-d/2)}{\Gamma(l) \Gamma(m)} \frac{\Gamma(d/2-l) \Gamma(d/2-m)}{\Gamma(d-l-m)},
\end{equation}
where  powers $l$ and $m$ greater than one may only arise from the terms proportional to $1-\xi$ in the gluon propagator (see Eq. \eqref{QQprop}). 
The only non-vanishing (i.e. non tadpole) integrals are 
%
\begin{subequations}\label{intregd}
\begin{eqnarray}
I^{11} & = & i \frac{(k^2)^{d/2-2}}{2^d\pi^{d/2}}
\frac{\Gamma \left(2-\frac{d}{2}\right) \Gamma \left(\frac{d}{2}-1\right)^2}{\Gamma (d-2)} \\
I^{12} & = & I^{21} = \frac{(3-d) } {k^2}  I^{11}\\ 
I^{22} & = & \frac{(3-d) (6-d) } {k^4}  I^{11}.
\end{eqnarray}
\end{subequations}

\begin{figure}
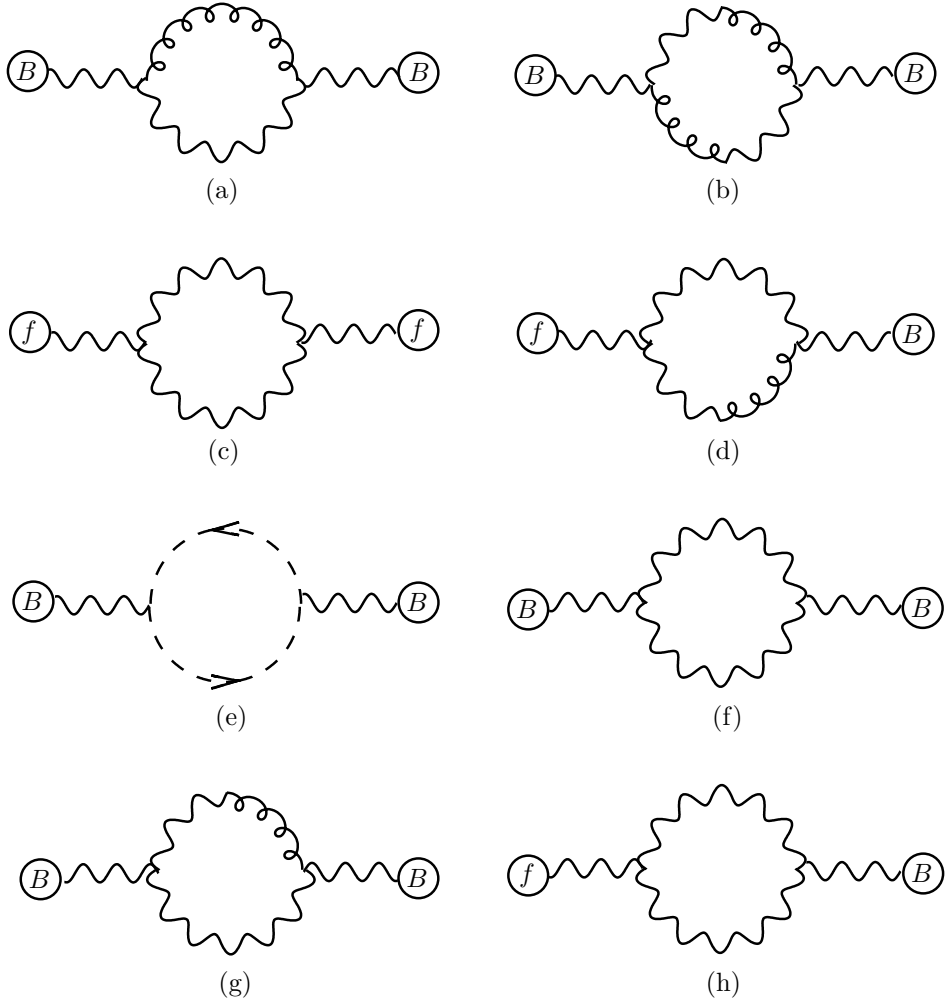

\begin{eqnarray}
\input{bfm1.pspdftex}  & \qquad \input{bfm2.pspdftex} 
 \nonumber 
\\
\nonumber 
\\
\input{bfm3.pspdftex}  & \qquad \input{bfm4.pspdftex}  
\nonumber 
\\
\nonumber 
\\
\input{bfm5.pspdftex}  & \qquad \input{bfm6.pspdftex}  
\nonumber
\nonumber 
\\
\nonumber 
\\
\input{bfm7.pspdftex}  & \qquad \input{bfm8.pspdftex}  
\nonumber
\end{eqnarray}
\caption{One-loop contributions to $\langle BB \rangle$.}
\label{fig1}
\end{figure}

Implementing the above described procedure as a straightforward computer algebra code, we readily 
obtain the following exact results for $C^{I}_1$ and $C^{I}_2$ 
\begin{subequations}
\begin{eqnarray}
  C^{\mbox{a}}_1 & = &
  \left[\frac{1}{4} (d-2) \xi +\frac{(2-d) (d-2)}{4 (d-1)}    \right]
  I^{11};\;\;\;\;\;\;
C^{\mbox{a}}_2  =   \frac{d-2}{4} I^{11}
\end{eqnarray}
\begin{eqnarray}
C^{\mbox{b}}_1 & = &  \frac 1 4  I^{11} ;\;\;\;\;\;\;
C^{\mbox{b}}_2 =  \frac{1-d}{4} I^{11}
\end{eqnarray}
\begin{eqnarray}
  C^{\mbox{c}}_1 & = &  \left[ \frac{1}{8} (d-4) \xi ^2-\frac{\xi }{2}-\frac{d}{8}  \right]  
  I^{11} ;\;\;\;\;\;\;
C^{\mbox{c}}_2 =  0
\end{eqnarray}
\begin{eqnarray}
  C^{\mbox{d}}_1 & = &  \left[      \frac{1}{2} (d-3) \xi + \frac{1-d}{2} \right]   
  I^{11} ;\;\;\;\;\;\;
C^{\mbox{d}}_2 =  0 
\end{eqnarray}
\begin{eqnarray}
C^{\mbox{e}}_1 & = &  \frac{1}{1-d}  I^{11} ;\;\;\;\;\;\;
C^{\mbox{e}}_2 =  0
\end{eqnarray}
\begin{eqnarray}
C^{\mbox{f}}_1 & = &  \frac{1}{4 (d-1)}  I^{11} ;\;\;\;\;\;\;
C^{\mbox{f}}_2  =   -\frac 1 4 I^{11}
\end{eqnarray}
\begin{eqnarray}
C^{\mbox{g}}_1 & = &  \left[\frac{1}{2 (d-1)}-1\right] I^{11} ;\;\;\;\;\;\;
C^{\mbox{g}}_2 =  \frac 1 2  I^{11}
\end{eqnarray}
\begin{eqnarray}
C^{\mbox{h}}_1 & = &  -\frac{1}{2} (\xi +1)  I^{11} ;\;\;\;\;\;\;
C^{\mbox{h}}_2 =  0 
\end{eqnarray}
\end{subequations}
Adding all the diagrams, we obtain the following {\it transverse} result for the one-loop contribution to $\langle BB \rangle$ 
\be\label{c9}
\Pi^{ab} _{\mu\nu} = N g^2 \delta^{ab}
\left[\frac{d-4}{8}  \xi ^2 + \frac{3(d-4)}{4}  \xi + \frac{1}{2-2 d}-\frac{7 d}{8}\right] I^{11} \left(k_\mu k_\nu - k^2\eta_{\mu\nu}\right) .
\ee

Finally, using $d=4-2\epsilon$ we obtain the following contribution for the UV pole
($I^{11} \approx {1}/(16 \pi^2 \epsilon )$)
\be\label{eqcla}
\left.\Pi_{\mu\nu}^{ab}(k)\right|_{UV} = -\frac{11}{3} i \delta^{ab} \frac{N g^2}{16\pi^2\epsilon}\left(k_\mu k_\nu - k^2\eta_{\mu\nu}\right). 
\ee
It is also interesting the note that graphs (c), (d) and (h), Fig. 1, which contains the linear part of 
$f^a_{\mu\nu}$, give the following UV gauge dependent contribution
\be\label{eqcl}
-\left(\frac{\xi }{2}+\frac{5}{2}\right)
 i \delta^{ab} \frac{N g^2}{16\pi^2\epsilon}\left(k_\mu k_\nu - k^2\eta_{\mu\nu}\right). 
\ee
This result shows that the term $Q_\mu\wedge Q_\nu \cdot f^{\mu\nu}(B)$
in Eq. \eqref{eq14} is indeed necessary 
in order to have a consistent gauge independent result, such as Eq. \eqref{eqcla}.
Since this contribution has been induced by the subtraction
from ${\cal L}_{YM}(B+Q)$,  in \eqref{eq9}, of the terms which 
are linear in $Q$, this calculation
of $\langle BB \rangle$ provides an explicit example of the consistency 
of the subtraction prescription. 

\section{Renormalization of the background field ${\cal F}_{\mu\nu}$}

The relevant Feynman graphs
(the ones that do not vanish upon using dimensional regularization)
contributing to the renormalization
of ${\cal F}_{\mu\nu}$ are shown in Fig. 2. 

\begin{figure}
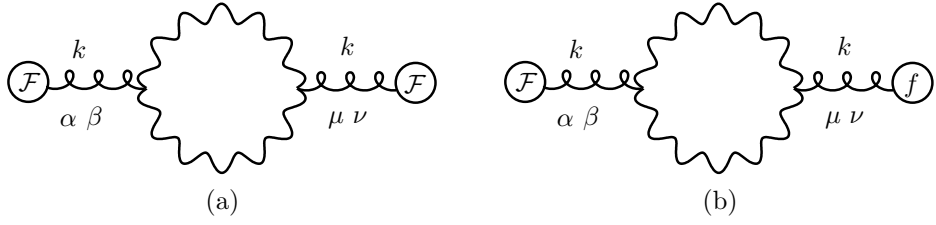

\begin{eqnarray}
\input{appendixDa.pspdftex}  & \qquad \input{appendixDb.pspdftex} 
 \nonumber 
\end{eqnarray}
\caption{One-loop contributions to $\langle {\cal F}{\cal F} \rangle$
  and $\langle {\cal F}f \rangle$.}
\end{figure}

Using the Feynman rules given in the Appendix B, together
with the result shown in Eq. \eqref{eq14}, one obtains for
the $\langle {\cal F}{\cal F}\rangle$ self-energy, the expression
\be\label{d1}
\Pi^{{\cal F} {\cal F}}_{\alpha\beta,\mu\nu} = -\frac{Ng^2}{16\pi^2\epsilon}
\frac{\xi+1}{2}\left(\eta_{\alpha\mu}\eta_{\beta\nu}-
\eta_{\alpha\nu}\eta_{\beta\mu}\right). 
\ee
Similarly, one gets for the mixed
$\langle {\cal F}{f}\rangle$ self-energy, the result
\be\label{d2}
\Pi^{{\cal F} f}_{\alpha\beta,\mu\nu} = -\frac{Ng^2}{16\pi^2\epsilon}
\frac{\xi+1}{2}\left(\eta_{\alpha\mu}\eta_{\beta\nu}-
\eta_{\alpha\nu}\eta_{\beta\mu}\right). 
\ee

Using the  above results, one finds from Eq. \eqref{eq44a}
that the counter-terms involved in the renormalization of
${\cal F}_{\mu\nu}$ are given by
\be
Z^{1/2}_{{\cal F}} = 1 + \frac{Ng^2}{16\pi^2\epsilon}
\frac{\xi+1}{4};\;\;\;\;
Z_{{\cal F}f} =  \frac{Ng^2}{16\pi^2\epsilon}
\frac{\xi+1}{2}.
\ee
Thus, the renormalization of the background field ${\cal F}_{\mu\nu}$
involves a rescaling as well as a mixing of the field.

\newpage


\end{document}